\documentclass[aps,prb,twocolumn,amssymb,amsmath,floatfix]{revtex4}

\usepackage{graphicx}
\usepackage{color}
\usepackage{amsmath}
\usepackage{ifsym}  %more symbols see...http://www.ctan.org/tex-archive/info/symbols/comprehensive/symbols-a4.pdf
\usepackage{amssymb}    %this allows \square etc... see http://www.artofproblemsolving.com/LaTeX/AoPS_L_GuideSym.php

\newcommand{\degC}{$^{\circ}$C}
\newcommand{\degrees}{$^{\circ}$}

\makeatletter
\makeatletter \renewcommand{\fnum@figure}
{\figurename~\thefigure}
\makeatother

\renewcommand{\figurename}{Figure}

\begin{document}

%Title of paper
\title{Linear magnetoresistance in commercial n-type silicon due to inhomogeneous doping}

\author{Nicholas A. Porter\footnote{email:phynap@leeds.ac.uk}}
\affiliation{School of Physics \& Astronomy, University of Leeds, Leeds LS2 9JT, United Kingdom}

\author{Christopher H. Marrows\footnote{email:c.h.marrows@leeds.ac.uk}}
\affiliation{School of Physics \& Astronomy, University of Leeds, Leeds LS2 9JT, United Kingdom}

\date{\today}

%\maketitle must follow title, authors, abstract, \pacs, and \keywords
\maketitle

%Nature Materials bold opening paragraph

\textbf{Free electron theory tells us that resistivity is independent of magnetic field\cite{Ashcroft1976}. In fact, most observations match the semiclassical prediction of a magnetoresistance that is quadratic at low fields before saturating\cite{PippardBook1989}. However, a non-saturating linear magnetoresistance has been observed in exotic semiconductors such as silver chalcogenides\cite{Xu1997,Husmann2002}, lightly-doped InSb\cite{HuNat2008}, N-doped InAs\cite{patane2009}, MnAs-GaAs composites\cite{Johnson2010}, PrFeAsO\cite{bhoi2011}, and epitaxial graphene\cite{friedman2010}. Here we report the observation of a large linear magnetoresistance in the ohmic regime in commonplace commercial n-type silicon wafer. It is well-described by a classical model of spatially fluctuating donor densities\cite{Herring1960}, and may be amplified by altering the aspect ratio of the sample to enhance current-jetting: increasing the width tenfold increased the magnetoresistance at 8 T from 445 \% to 4707 \% at 35 K. This physical picture may well offer insights into the large magnetoresistances recently observed in n-type\cite{DelmoNature2009} and p-type\cite{Schoonus2008} Si in the non-ohmic regime.}

The conventional theory of magnetoresistance (MR) in metals and semiconductors relies upon a distribution of scattering times amongst the conducting carriers that cannot be compensated by a unique Hall field\cite{PippardBook1989}. For a material with a closed free electron Fermi surface
(FS) and a principal charge carrier this leads to positive MR that is quadratic in weak magnetic fields, $B$, and saturates in strong magnetic fields according to:
\begin{equation}
\frac{\Delta\rho}{\rho} = \frac{\rho(B) - \rho(0)}{\rho(0)} =
\frac{A B^{2}}{1 + C B^{2}},\label{eqn:positiveMR}
\end{equation}
where $\rho$ is the resistivity. In the denominator of equation \ref{eqn:positiveMR} it can be shown\cite{sommerfeld1931} that $C B^{2} = (l_{\mathrm{MF}}/r_{\mathrm{c}})^{2}=(\omega_{c}\tau)^{2}$, where $\omega_{c}$ is the cyclotron frequency, $\tau$ the mean free time between scattering events, $l_{\mathrm{MF}}$ the mean free path and $r_{\mathrm{c}}$ the cyclotron radius. In the strong field limit
($\omega_{c}\tau \gg 1$), electrons complete several orbits before scattering. In the weak field limit ($\omega_{c}\tau \ll 1$), only a fraction of these orbits are completed, and so $\Delta\rho/\rho \sim A B^{2}$, depending predominantly upon the anisotropy in relaxation times and cyclotron masses of the carriers\cite{sommerfeld1931}. Other galvanomagnetic effects are present in magnetically ordered materials\cite{ohandleybook}.

In the past decade some materials have emerged that break these previously established trends. Narrow gap semiconductors such as iron diantinomide\cite{Petrovic2003}, the silver chalcogenides\cite{Xu1997}, indium antimonide\cite{HuNat2008}, and nitrogen-doped indium arsenide\cite{patane2009} have shown unexpectedly high MR over large temperature ranges. In each case, the MR was non-saturating, approximately linear at high fields, and was enhanced by strong disorder arising from impurity substitution\cite{Hu2008MR,Manyala2000} and spatial inhomogeneity of the sample stoichiometry\cite{Xu1997,Hu2005,patane2009}. In the last few years numerical modelling of large scale inhomogeneous inclusions using classical models has been used to describe this linear MR\cite{Parish2003,Parish2005}.

Here we have measured the MR in n-type commercial Si wafer, which we attribute to a more subtle mechanism
due to the weak disorder from spatially fluctuating conductivity that arises statistically from the random donor distribution\cite{Herring1960}. Si:P wafers with $3.6\pm0.3$ $\Omega$cm room temperature resistivity and dopant density $(1.4 \pm 0.1) \times 10^{15}$ cm$^{-3}$ were used. The net carrier density, $\langle n \rangle$, and an absolute value of the carrier mobility, $\mu$, were obtained as a function of temperature, $T$, using the van der Pauw method on square wafers (see supplementary figure 1). The MR was measured using four in-line contacts on a sample with the following dimensions: width, $w = 1.8 \pm 0.1$ mm; length, $l = 5.5 \pm 0.5$ mm; thickness, $t = 530 \pm 50$ $\mu$m; and voltage probe separation, $s = 2.0 \pm 0.4$ mm. Indium ohmic contacts provided a four wire current-voltage $I$-$V$ characteristic that was ohmic at low bias throughout all temperatures. Measurements were performed in three orientations of field with respect to current density $\underline{j}$: transverse ($\underline{j} \perp \underline{B}$) with the magnetic field out of the sample plane; perpendicular ($\underline{j} \perp \underline{B}$) with the magnetic field in plane; and longitudinal ($\underline{j} \parallel \underline{B}$).

\begin{figure}[t]
\begin{center}
\includegraphics[width=8.5cm]{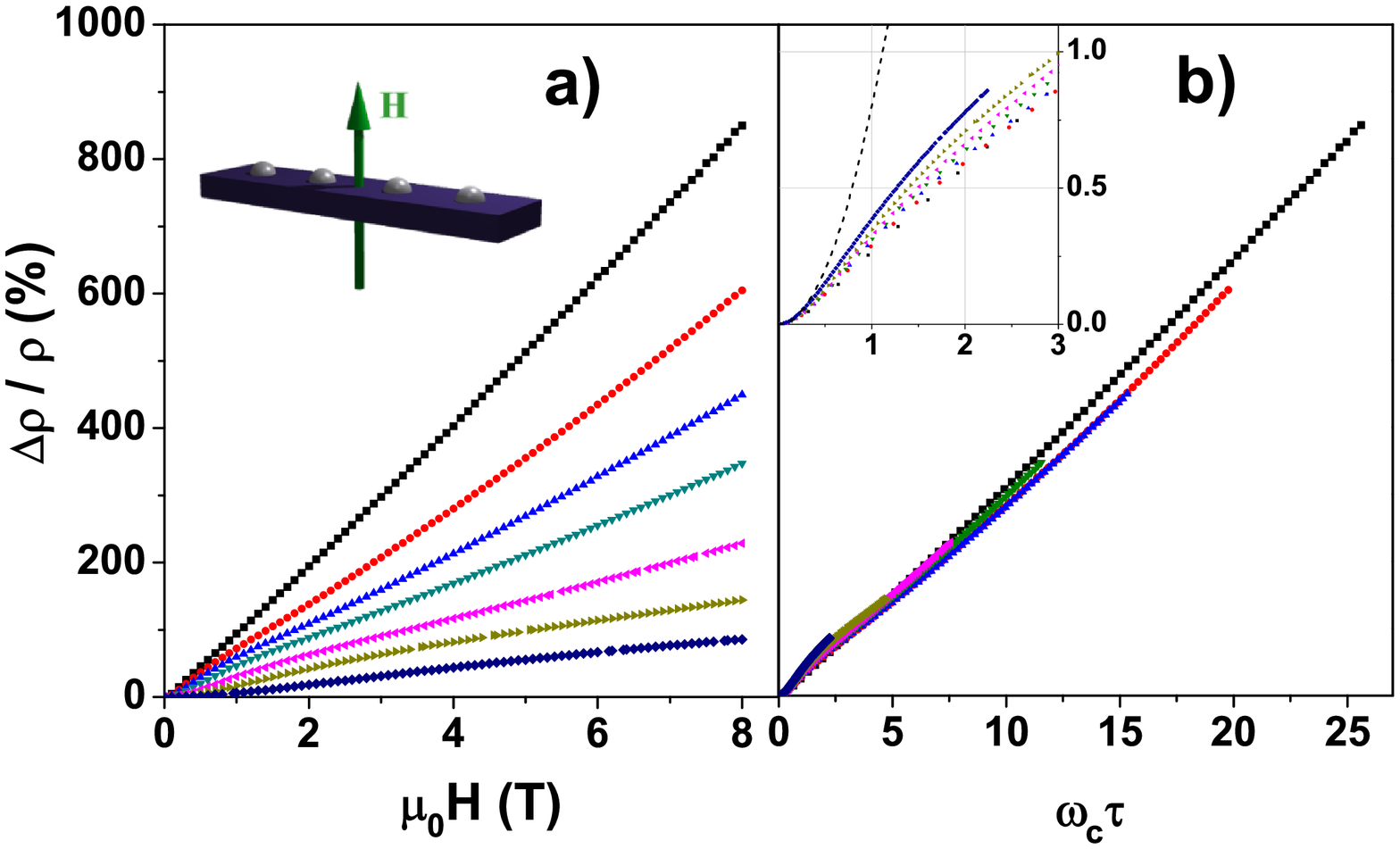}
\end{center}
\caption{Transverse MR at: 30 ($\blacksquare$), 40 (\textcolor[rgb]{1,0,0}{$\bullet$}), 50 (\textcolor[rgb]{0,0,1}{$\blacktriangle$}), 60 (\textcolor[rgb]{0,0.5,0.5}{$\blacktriangledown$}), 80
(\textcolor[rgb]{1,0,1}{$\blacktriangleleft$}), 120 (\textcolor[rgb]{0.5,0.5,0}{$\blacktriangleright$}) and 200 K (\textcolor[rgb]{0,0,0.5}{$\blacklozenge$}) in transverse magnetic fields. a) MR increased
monotonically with decreasing temperature. In the strong magnetic field condition MR was non-saturating and approximately linear. b) MR at all temperatures reduced down to one unique curve provided $\mu B =
\omega_{c} \tau$ conforming to Kohler's rule. The data is magnified in the inset to demonstrate the weak to strong field crossover at $\omega_{c}\tau \sim 1$ and universal scaling of the weak field MR. A quadratic law is plotted at weak fields, $\omega_{c}\tau \ll 1$ (dashed line)}\label{fig:transMR}
\end{figure}

The transverse MR in the ohmic regime in figure \ref{fig:transMR}a was approximately linear up to high magnetic fields and showed no indication of saturation in 8 T fields at temperatures from 30 -- 200 K, and is much larger than at room temperature where the system barely enters the high-field regime\cite{porter2011}. The high field gradient of this plot is known as the Kohler slope (KS). The data were reduced to a Kohler plot of $\Delta\rho/\rho$ versus $\omega_{c}\tau$, shown in \ref{fig:transMR}b. In weak and strong fields the MR scaled onto a unique curve according Kohler's law; $\Delta\rho/\rho = f(\omega_{c}\tau)$, where $f$ is a scaling function, using the relationship $\omega_{c}\tau = \mu B$. Longitudinal measurements did not obey Kohler's law but possessed a variable KS as a function of temperature as shown in the supplementary figures 2 and 3. In weak fields (shown in the inset of figure \ref{fig:transMR}b) the departure from quadratic behaviour (shown as dashed line) occurred for less than half a completed cyclotron orbit, while linear MR was satisfied in strong fields for $\omega_{c}\tau \gtrsim 2$.

Linear MR can arise in polycrystalline material with an open FS resulting from the averaging of saturating MR from closed orbits and non-saturating MR from open orbits\cite{Lifshitz1956}. Such effects can also be caused by quantum linear magnetoresistance (QLMR)\cite{Abrikosov2003,HuNat2008}. The former can be disregarded for the single crystals used here and the closed silicon conduction band energy surfaces. The latter requires a
very small $n$ and effective mass, $m^{\star}$ which generally limits it to metals and semi-metals. As $m^{\star}\sim 0.4 m_{e}$ in Si:P\cite{Ashcroft1976}, a field of 10 T would not produce QLMR
above 30 K\cite{Abrikosov2003} since the thermal energy would exceed the separation of the Landau levels, ruling out this mechanism here.

Aside from these possible causes, Pippard proposed four distinct mechanisms that could lead to strong field, non saturating linear MR in crystals\cite{PippardBook1989}: small angle scattering, magnetic breakdown, voids, and inhomogeneities.

The influence of small angle scattering is limited at higher temperatures, where electrons are scattered through larger angles. In Si:P at 100 K, where a large positive MR is observed, the most effective phonons would scatter at $\sim 30$\degrees\ suggesting that small angle scattering is not applicable\cite{PippardBook1989}. Equally, magnetic breakdown from tunnelling between orbits in momentum-space is unlikely as the FS of the conduction band in Si lies far from the Brillouin zone boundary.

MR arising from spherical voids with a volume fraction $F$ scales as $0.486 F \omega_{c}\tau$ in a transverse field and $0.637 F \omega_{c}\tau$ in a longitudinal field\cite{PippardBook1989}. As the magnitude of the transverse MR shown in figure \ref{fig:transMR}a is much greater than the longitudinal MR (see supplementary figure 2) the observed MR does not agree with this hypothesis. In figure \ref{fig:transMR}b, with the aspect ratio $l/w \sim 3$, the KS was $0.30 \pm 0.01$, implying a void volume fraction $F=0.62 \pm 0.02$ in our single crystal Si wafer, which is utterly implausible.

Thus, linear MR from inhomogeneities remains as the most probable cause of enhanced MR in strong fields. The limit of strong disorder has been used to explain the linear MR in the silver chalcogenides\cite{Parish2005}. However, silicon growth techniques aim to produce uniformly doped wafers and thus the disorder is likely to be weak. The suspicion that fluctuation in dopant density of pulled crystals may influence the conduction of semiconducting single crystals was the original motivation for Herring's paper in 1960\cite{Herring1960}, although there was little direct experimental evidence available at that time. The spatial variations in Hall coefficient that these defects give rise to were predicted to prevent saturation of the transverse MR whilst having no effect on the saturating longitudinal MR. For instance, in the non-degenerate weakly doped sample discussed here, the average impurity separation is $\sim 90$ nm. The mean free path at 200 K, $l_{\mathrm{MF}} = ( \mu / e )( 3 m^{\star}k_{\mathrm{B}} T )^{1/2}$, where $e$ is the electron charge and $k_{\mathrm{B}}$ is Boltzmann's constant, would be $\sim 92$ nm\cite{Weber1991}. As the electrons will scatter on a lengthscale comparable to the impurity separation they can be susceptible to fluctuations in the concentration of ionised donors, $N_{\rm i}$, on lengthscales greater than their mean free path. Herring suggested that the fractional variance (FV) of the number of ionised donors $\langle ( N_{\rm i} - \langle N_{\rm i}\rangle)^{2} \rangle/ \langle N_{\rm i} \rangle^{2}$  was the relevant measure of the fluctuation magnitude. If this value was significant over lengthscales greater than a suitably defined effective mean free path, $l_{\mathrm{eff}}$, and the Debye screening length, $\lambda_{\mathrm{D}}$, then the inhomogeneity that results from the distribution of random impurities would create uncompensated Hall fields that contribute significantly to a linear MR\cite{Herring1960}.

The fractional variance should hence be limited on a lengthscale $\lambda_{D}=\sqrt{\epsilon k_{\mathrm{B}} T / (\langle n \rangle e^{2})}$, where $\epsilon$ is the Si:P permittivity, in lightly doped samples, and by $l_{\mathrm{eff}}=(2 / (\omega_{\mathrm{c}}\tau)^{2})^{-1/3}l_{\mathrm{MF}}$ in highly doped samples. Isotropic, saturating MR such as in equation \ref{eqn:positiveMR} is assumed in the case of homogeneous doping. When donor density fluctuations are introduced, the effective MR for transverse isotropic fluctuations in $N_{\rm i}$ can be written as\cite{Herring1960}:
\begin{equation}
\frac{\Delta \rho_{\mathrm{eff}}}{\rho_{\mathrm{eff}}} \sim \frac{\Delta \rho}{\rho} +
\frac{B}{n e \rho(0)}\zeta \frac{\langle ( N_{\rm i} - \langle N_{\rm i} \rangle)^{2} \rangle}{\langle N_{\rm i} \rangle^{2}},\label{eqn:HerringMR}
\end{equation}
where $\zeta$ has a numerical value that depends upon the conductivity anisotropy. In the isotropic case relevant here, $\zeta = \pi /4$.

\begin{figure}[t]
\begin{center}
\includegraphics[width=8cm]{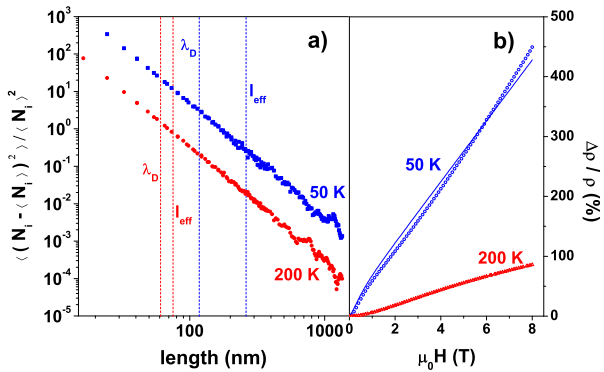}
\caption{Interpretation of the observed MR within the Herring model. (a) Simulation of the impurity concentration fluctuations of a cubic crystal of $10^{12}$ atoms on different lengthscales. The vertical lines mark the Debye screening length, $\lambda_{\mathrm{D}}$, and effective mean free path, $l_{\mathrm{eff}}$, at 50 K and 200 K. These parameters provide an upper limit for the FV to which the electron will be sensitive. For this impurity concentration the FV is limited by $l_{\mathrm{eff}}$
at both temperatures but still remains large. b) Fits (lines) to the transverse MR (symbols) at 50 and 200 K according to the simplified model of Herring in equation \ref{eqn:HerringMR}.}\label{fig:HerringModel}
\end{center}
\end{figure}

To estimate the FV of a random distribution of impurities over the relevant lengthscales, we simulated a cubic crystal consisting of $10^{12}$ atoms, substituting random host atoms for ionised impurities. The FV of the impurities probed over different lengthscales is shown in figure \ref{fig:HerringModel}a. The average density of ionised impurities at 50 and 200 K was taken from supplementary figure 1 assuming that $N_{\rm i} \approx n$ in ohmic conduction. Included in figure \ref{fig:HerringModel}a are the lower limits on the FV at each temperature given by the lengthscales $\lambda_{\mathrm{D}}$ and $l_{\mathrm{eff}}$, calculated for $\omega_{\mathrm{c}}\tau = 2$. These limitations suggest that the FV should be no more than 0.27 at 50 K and 0.77 at 200 K.

To see if these values of the FV in impurity density are sufficient to produce the observed linear
MR, fits to equation \ref{eqn:HerringMR} were performed using the measured $\mu$, $\rho(0)$, and $\langle n \rangle$ values with only the FV and $A$ as fitting parameters. The fits are shown in figure
\ref{fig:HerringModel}b, showing that a FV of 0.34 is required at 50 K and 0.21 at 200 K. At 50 K, it is expected that the value for $\omega_{\mathrm{c}}\tau$ for which these conditions are satisfied should be 2; from our fits, the value for $\omega_{\mathrm{c}}\tau = 2.25$ is in reasonable agreement. Overall, the classical Herring model provides a very satisfactory explanation for our observations.

%%%%%%%%%%%%%%%%GEOMETRICAL ENHANCEMENT%%%%%%%%%%%%%%%%%

This non-saturating linear MR of the Si:P can be enhanced by classically understood methods, by altering the aspect ratio of the wafers, which causes current jetting. When attempting to discern the physical MR of semiconductors, such geometrical enhancement is to be avoided, as above. However, it can be used to enhance the response of devices in magnetic fields. Solin et al. have shown that the geometrical enhancement in
high mobility semiconductors in certain geometries such as Corbino disks can be used for practical applications such as read heads for high-density recording\cite{Solin2000,Solin2002}.

\begin{figure}[t]
\includegraphics[width=8.5cm]{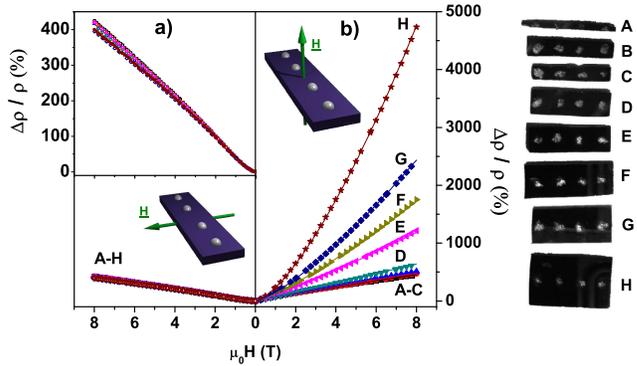}
\caption{Geometrical enhancement MR, measured in perpendicular (a) and transverse (b) field orientations at 35 K shown mirrored to one another using the right hand axis to encourage comparison. Wafers A-H are shown with respective values of $\alpha = l / w$: $10  \pm 1  $ (A,$\blacksquare$), $4.5 \pm 0.6$ (B,\textcolor[rgb]{1,0,0}{$\bullet$}), $3.7 \pm 0.5$ (C,\textcolor[rgb]{0,0,1}{$\blacktriangle$}), $2.6 \pm 0.3$ (D,\textcolor[rgb]{0,0.5,0.5}{$\blacktriangledown$}), $2.3 \pm 0.2$ (E,\textcolor[rgb]{1,0,1}{$\blacktriangleleft$}), $1.9 \pm 0.3$ (F,\textcolor[rgb]{0.5,0.5,0}{$\blacktriangleright$}), $1.6 \pm 0.2$
(G,\textcolor[rgb]{0,0,0.5}{$\blacklozenge$}), $1.1 \pm 0.2$
(H,\textcolor[rgb]{0.5,0,0}{$\bigstar$}).}\label{fig:GeomTrans}
\end{figure}

We prepared nine further samples with increasing width, $w$, with indium contacts positioned at a separation $l \sim$ 10 mm between current leads and $s \sim$ 3 mm between voltage probes. Photographs of these samples, labelled A-H, are shown in figure \ref{fig:GeomTrans} with the dimensions, measured using vernier calipers, given in the caption. Perpendicular and transverse MR at 35 K of samples A to H are shown in figures \ref{fig:GeomTrans}a and \ref{fig:GeomTrans}b respectively. In the perpendicular orientation, where the sample projection in the field direction does not change, the variation in MR between the samples is very small. In the transverse field shown in figure \ref{fig:GeomTrans}b the projection of the sample in the magnetic field direction varied significantly. In this orientation, increasing the width of the samples dramatically enhanced the MR of the films. Sample A, the narrowest sample, reached a maximum $\Delta\rho / \rho = 445$ \% at 8 T. The widest sample, H, had a maximum $\Delta\rho / \rho = 4707$ \%, over an order of
magnitude larger.

Geometrical enhancement of transverse MR has been calculated for rectangular plates of homogeneous conductors with electrodes spanning the sample ends\cite{PopovicBook2004,Heremans1993}:our geometry differs slightly in that we have point-like contacts, but the same general conclusions hold. For the narrowest sample A, the net current flow would be approximately homogeneous providing a good indication of the physical MR intrinsic to the material. In sample H, the widest sample, current flow would be inhomogeneous with significant current deflection perpendicular to the magnetic field near to the electrodes, increasing the voltage detected there and hence the measured resistance.

%%%%%%%%%%%%%%%%DISCUSSION%%%%%%%%%%%%%%%%%
Hence, we conclude that the large non-saturating MR we observe here relies upon the distribution of ionised donors being sufficiently disordered that current distortions arise from the local fluctuating Hall fields. The donor distribution in this study is random, defined by a three-dimensional Poisson distribution, but with modern lithographic and ion implantation techniques, the local density could be tailored to maximise this effect. Alternatively, for a thin sample, lithography techniques could produce an array of top gates that could tune the local density of ionised donors to provide strong disorder by locally modifying the Hall constant, providing a MR response that can be tailored with an applied electric field.

If the carrier concentration is decreased, as is the case for very lightly doped silicon, the linear MR from the FV in equation \ref{eqn:HerringMR} is limited by the Debye electron screening length. This limitation from screening could perhaps be overcome by the large electric fields resulting from  the concentration of
conduction electrons. As $\lambda_{\mathrm{D}} \propto \langle n \rangle^{-1/2}$, in the non-quasineutral Mott-Gurney regime, $\lambda_{\mathrm{D}}$ would be reduced without changing the FV of the ionised impurities. As the FV would remain large, with the reduced limitation from $\lambda_{\mathrm{D}}$, a large FV may produce a large MR even in intrinsic samples. This could help to explain the large MR in high electric fields reported for intrinsic silicon (i-Si)\cite{DelmoNature2009,Schoonus2009,Ciccarelli2010}. For instance, with typically $\langle N_{\rm i} \rangle \sim 10^{12}$ cm$^{3}$ at 300 K, $\langle ( N_{\rm i} - \langle N_{\rm i}\rangle)^{2} \rangle / \langle N_{\rm i} \rangle^{2} \sim 0.01$. By increasing the source-drain current to inject more carriers such that $\langle n_{\mathrm{inj}} \rangle \sim 10 \langle n \rangle$ this value would increase to $\sim 0.5$. It is possible that the significant enhancement that could arise in both Mott-Gurney regime and during electrical breakdown could increase the MR to produce the large measured values\cite{DelmoNature2009,Schoonus2009}.

\section*{Methods}

\footnotesize{
In all wafers the $\langle001\rangle$ crystal vector was normal to the surface. High angle XRD verified that the wafers were cleaved along $\langle110\rangle$ orientations such that the current, $\underline{j}$, was parallel to $\langle110\rangle$ in all measurements.

Contacts were made to the wafer surface by mechanical cleavage through the native oxide with a diamond scribe followed by impression of indium. The contacts were then annealed at 350 \degC\ for 10 minutes. When the contacts were made, aluminium wire was attached using a wedge bonder from the sample holder to the indium.
A 3 $\times$ 3 mm$^{2}$ square wafer was used for van der Pauw measurements using magnetic fields of $\pm$1.5 T to ascertain Hall voltages. All magnetoresistance measurements were performed in a gas flow cryostat using a four probe in-line DC technique at low bias in the ohmic regime, with the current bias adjusted to maintain roughly 5 mV between the voltage probes.
}

\bibliographystyle{naturemag}
%% Create the reference section using BibTeX:
%%\bibliography{basename of .bib file}
%\bibliography{LinearMRSi}

\begin{thebibliography}{10}
\expandafter\ifx\csname url\endcsname\relax
  \def\url#1{\texttt{#1}}\fi
\expandafter\ifx\csname urlprefix\endcsname\relax\def\urlprefix{URL }\fi
\providecommand{\bibinfo}[2]{#2}
\providecommand{\eprint}[2][]{\url{#2}}

\bibitem{Ashcroft1976}
\bibinfo{author}{Ashcroft, N.~W.} \& \bibinfo{author}{Mermin, N.~D.}
\newblock \emph{\bibinfo{title}{Solid State Physics}}
  (\bibinfo{publisher}{Harcourt Brace}, \bibinfo{address}{Fort Worth},
  \bibinfo{year}{1976}).

\bibitem{PippardBook1989}
\bibinfo{author}{Pippard, A.~B.}
\newblock \emph{\bibinfo{title}{Magnetoresistance in Metals}}
  (\bibinfo{publisher}{Cambridge University Press},
  \bibinfo{address}{Cambridge}, \bibinfo{year}{1989}).

\bibitem{Xu1997}
\bibinfo{author}{Xu, R.} \emph{et~al.}
\newblock \bibinfo{title}{Large magnetoresistance in non-magnetic silver
  chalcogenides}.
\newblock \emph{\bibinfo{journal}{Nature}} \textbf{\bibinfo{volume}{390}},
  \bibinfo{pages}{57--60} (\bibinfo{year}{1997}).

\bibitem{Husmann2002}
\bibinfo{author}{Husmann, A.} \emph{et~al.}
\newblock \bibinfo{title}{Megagauss sensors}.
\newblock \emph{\bibinfo{journal}{Nature}} \textbf{\bibinfo{volume}{417}},
  \bibinfo{pages}{421} (\bibinfo{year}{2002}).

\bibitem{HuNat2008}
\bibinfo{author}{Hu, J.} \& \bibinfo{author}{Rosenbaum, T.~F.}
\newblock \bibinfo{title}{Classical and quantum routes to linear
  magnetoresistance}.
\newblock \emph{\bibinfo{journal}{Nature Mater.}} \textbf{\bibinfo{volume}{7}},
  \bibinfo{pages}{697--700} (\bibinfo{year}{2008}).

\bibitem{patane2009}
\bibinfo{author}{Patan\`{e}, A.} \emph{et~al.}
\newblock \bibinfo{title}{Effect of low nitrogen concentrations on the
  electronic properties of {InAs}$_{1-x}${N}$_x$}.
\newblock \emph{\bibinfo{journal}{Phys. Rev. B}} \textbf{\bibinfo{volume}{80}},
  \bibinfo{pages}{115207} (\bibinfo{year}{2009}).

\bibitem{Johnson2010}
\bibinfo{author}{Johnson, H.~G.}, \bibinfo{author}{Bennett, S.~P.},
  \bibinfo{author}{Barua, R.}, \bibinfo{author}{Lewis, L.~H.} \&
  \bibinfo{author}{Heiman, D.}
\newblock \bibinfo{title}{Universal properties of linear magnetoresistance in
  strongly disordered {MnAs}-{GaAs} composite semiconductors}.
\newblock \emph{\bibinfo{journal}{Phys. Rev. B}} \textbf{\bibinfo{volume}{82}},
  \bibinfo{pages}{085202} (\bibinfo{year}{2010}).

\bibitem{bhoi2011}
\bibinfo{author}{Bhoi, D.}, \bibinfo{author}{Mandal, P.},
  \bibinfo{author}{Choudhury, P.}, \bibinfo{author}{Pandya, S.} \&
  \bibinfo{author}{Ganesan, V.}
\newblock \bibinfo{title}{Quantum magnetoresistance of the {PrFeAsO}
  oxypnictide}.
\newblock \emph{\bibinfo{journal}{Appl. Phys. Lett.}}
  \textbf{\bibinfo{volume}{98}}, \bibinfo{pages}{172105}
  (\bibinfo{year}{2011}).

\bibitem{friedman2010}
\bibinfo{author}{Friedman, A.~L.} \emph{et~al.}
\newblock \bibinfo{title}{Quantum linear magnetoresistance in multilayer
  epitaxial graphene}.
\newblock \emph{\bibinfo{journal}{Nano. Lett.}} \textbf{\bibinfo{volume}{10}},
  \bibinfo{pages}{3962} (\bibinfo{year}{2010}).

\bibitem{Herring1960}
\bibinfo{author}{Herring, C.}
\newblock \bibinfo{title}{Effect of random inhomogeneities on electrical and
  galvanomagnetic measurements}.
\newblock \emph{\bibinfo{journal}{J. Appl. Phys.}}
  \textbf{\bibinfo{volume}{31}}, \bibinfo{pages}{1939--1953}
  (\bibinfo{year}{1960}).

\bibitem{DelmoNature2009}
\bibinfo{author}{Delmo, M.~P.}, \bibinfo{author}{Yamamoto, S.},
  \bibinfo{author}{Kasai, S.}, \bibinfo{author}{Ono, T.} \&
  \bibinfo{author}{Kobayashi, K.}
\newblock \bibinfo{title}{Large positive magnetoresistive effect in silicon
  induced by the space-charge effect}.
\newblock \emph{\bibinfo{journal}{Nature}} \textbf{\bibinfo{volume}{457}},
  \bibinfo{pages}{1112--1115} (\bibinfo{year}{2009}).

\bibitem{Schoonus2008}
\bibinfo{author}{Schoonus, J. J. H.~M.}, \bibinfo{author}{Bloom, F.~L.},
  \bibinfo{author}{Wagemans, W.}, \bibinfo{author}{Swagten, H. J.~M.} \&
  \bibinfo{author}{Koopmans, B.}
\newblock \bibinfo{title}{Extremely large magnetoresistance in boron-doped
  silicon}.
\newblock \emph{\bibinfo{journal}{Phys. Rev. Lett.}}
  \textbf{\bibinfo{volume}{100}}, \bibinfo{pages}{127202}
  (\bibinfo{year}{2008}).

\bibitem{sommerfeld1931}
\bibinfo{author}{Sommerfeld, A.} \& \bibinfo{author}{Frank, N.~H.}
\newblock \bibinfo{title}{The statistical theory of thermoelectric, galvano-
  and thermomagnetic phenomena in metals}.
\newblock \emph{\bibinfo{journal}{Rev. Mod. Phys.}}
  \textbf{\bibinfo{volume}{3}}, \bibinfo{pages}{1--42} (\bibinfo{year}{1931}).

\bibitem{ohandleybook}
\bibinfo{author}{{O'Handley}, R.~C.}
\newblock \emph{\bibinfo{title}{Modern Magnetic Materials}}
  (\bibinfo{publisher}{Wiley}, \bibinfo{address}{New York},
  \bibinfo{year}{2000}).

\bibitem{Petrovic2003}
\bibinfo{author}{Petrovic, C.} \emph{et~al.}
\newblock \bibinfo{title}{Anisotropy and large magnetoresistance in the
  narrow-gap semiconductor {FeSb}$_{2}$}.
\newblock \emph{\bibinfo{journal}{Phys. Rev. B}} \textbf{\bibinfo{volume}{67}},
  \bibinfo{pages}{155205} (\bibinfo{year}{2003}).

\bibitem{Hu2008MR}
\bibinfo{author}{Hu, R.} \emph{et~al.}
\newblock \bibinfo{title}{Colossal positive magnetoresistance in a doped nearly
  magnetic semiconductor}.
\newblock \emph{\bibinfo{journal}{Phys. Rev. B}} \textbf{\bibinfo{volume}{77}},
  \bibinfo{pages}{085212} (\bibinfo{year}{2008}).

\bibitem{Manyala2000}
\bibinfo{author}{Manyala, N.} \emph{et~al.}
\newblock \bibinfo{title}{Magnetoresistance from quantum interference effects
  in ferromagnets}.
\newblock \emph{\bibinfo{journal}{Nature}} \textbf{\bibinfo{volume}{404}},
  \bibinfo{pages}{581} (\bibinfo{year}{2000}).

\bibitem{Hu2005}
\bibinfo{author}{Hu, J.}, \bibinfo{author}{Rosenbaum, T.~F.} \&
  \bibinfo{author}{Betts, J.~B.}
\newblock \bibinfo{title}{Current jets, disorder, and linear magnetoresistance
  in the silver chalcogenides}.
\newblock \emph{\bibinfo{journal}{Phys. Rev. Lett.}}
  \textbf{\bibinfo{volume}{95}}, \bibinfo{pages}{186603}
  (\bibinfo{year}{2005}).

\bibitem{Parish2003}
\bibinfo{author}{Parish, M.~M.} \& \bibinfo{author}{Littlewood, P.~B.}
\newblock \bibinfo{title}{Non-saturating magnetoresistance in heavily
  disordered semiconductors}.
\newblock \emph{\bibinfo{journal}{Nature}} \textbf{\bibinfo{volume}{426}},
  \bibinfo{pages}{162--165} (\bibinfo{year}{2003}).

\bibitem{Parish2005}
\bibinfo{author}{Parish, M.~M.} \& \bibinfo{author}{Littlewood, P.~B.}
\newblock \bibinfo{title}{Classical magnetotransport of inhomogeneous
  conductors}.
\newblock \emph{\bibinfo{journal}{Phys. Rev. B}} \textbf{\bibinfo{volume}{72}},
  \bibinfo{pages}{094417} (\bibinfo{year}{2005}).

\bibitem{porter2011}
\bibinfo{author}{Porter, N.~A.} \& \bibinfo{author}{Marrows, C.~H.}
\newblock \bibinfo{title}{Dependence of magnetoresistance on dopant density in
  phosphorous doped silicon}.
\newblock \emph{\bibinfo{journal}{J. Appl. Phys.}}
  \textbf{\bibinfo{volume}{109}}, \bibinfo{pages}{07C703}
  (\bibinfo{year}{2011}).

\bibitem{Lifshitz1956}
\bibinfo{author}{Lifshitz, I.~M.}, \bibinfo{author}{Azbel, M.~I.} \&
  \bibinfo{author}{Kaganov, M.~I.}
\newblock \bibinfo{title}{On the theory of galvanomagnetic effects in metals}.
\newblock \emph{\bibinfo{journal}{Sov. Phys. JETP}}
  \textbf{\bibinfo{volume}{3}}, \bibinfo{pages}{143--145}
  (\bibinfo{year}{1956}).

\bibitem{Abrikosov2003}
\bibinfo{author}{Abrikosov, A.~A.}
\newblock \bibinfo{title}{Quantum linear magnetoresistance; solution of an old
  mystery}.
\newblock \emph{\bibinfo{journal}{J. Phys. A: Math. Gen.}}
  \textbf{\bibinfo{volume}{36}}, \bibinfo{pages}{9119} (\bibinfo{year}{2003}).

\bibitem{Weber1991}
\bibinfo{author}{Weber, L.} \& \bibinfo{author}{Gmelin, E.}
\newblock \bibinfo{title}{Transport properties of silicon}.
\newblock \emph{\bibinfo{journal}{Appl. Phys. A}}
  \textbf{\bibinfo{volume}{53}}, \bibinfo{pages}{136--140}
  (\bibinfo{year}{1991}).

\bibitem{Solin2000}
\bibinfo{author}{Solin, S.~A.}, \bibinfo{author}{Thio, T.},
  \bibinfo{author}{Hines, D.~R.} \& \bibinfo{author}{Heremans, J.~J.}
\newblock \bibinfo{title}{Enhanced room-temperature geometric magnetoresistance
  in inhomogeneous narrow-gap semiconductors}.
\newblock \emph{\bibinfo{journal}{Science}} \textbf{\bibinfo{volume}{289}},
  \bibinfo{pages}{1530--1532} (\bibinfo{year}{2000}).

\bibitem{Solin2002}
\bibinfo{author}{Solin, S.~A.} \emph{et~al.}
\newblock \bibinfo{title}{Nonmagnetic semiconductors as read-head sensors for
  ultra-high-density magnetic recording}.
\newblock \emph{\bibinfo{journal}{Appl. Phys. Lett.}}
  \textbf{\bibinfo{volume}{80}}, \bibinfo{pages}{4012--4014}
  (\bibinfo{year}{2002}).

\bibitem{PopovicBook2004}
\bibinfo{author}{Popovic, R.~S.}
\newblock \emph{\bibinfo{title}{Hall Effect Devices}}
  (\bibinfo{publisher}{Institute of Physics Publishing},
  \bibinfo{address}{London}, \bibinfo{year}{2004}), \bibinfo{edition}{2nd} edn.

\bibitem{Heremans1993}
\bibinfo{author}{Heremans, J.}
\newblock \bibinfo{title}{Solid state magnetic field sensors and applications}.
\newblock \emph{\bibinfo{journal}{J. Phys. D Appl. Phys.}}
  \textbf{\bibinfo{volume}{26}}, \bibinfo{pages}{1149} (\bibinfo{year}{1993}).

\bibitem{Schoonus2009}
\bibinfo{author}{Schoonus, J. J. H.~M.}, \bibinfo{author}{Haazen, P. P.~J.},
  \bibinfo{author}{Swagten, H. J.~M.} \& \bibinfo{author}{Koopmans, B.}
\newblock \bibinfo{title}{Unravelling the mechanism of large room-temperature
  magnetoresistance in silicon}.
\newblock \emph{\bibinfo{journal}{J. Phys. D: Appl. Phys.}}
  \textbf{\bibinfo{volume}{42}}, \bibinfo{pages}{185011}
  (\bibinfo{year}{2009}).

\bibitem{Ciccarelli2010}
\bibinfo{author}{Ciccarelli, C.}, \bibinfo{author}{Park, B.~G.},
  \bibinfo{author}{Ogawa, S.}, \bibinfo{author}{Ferguson, A.~J.} \&
  \bibinfo{author}{Wunderlich, J.}
\newblock \bibinfo{title}{Gate controlled magnetoresistance in a silicon
  metal-oxide-semiconductor field-effect-transistor}.
\newblock \emph{\bibinfo{journal}{Appl. Phys. Lett.}}
  \textbf{\bibinfo{volume}{97}}, \bibinfo{pages}{082106}
  (\bibinfo{year}{2010}).

\end{thebibliography}

\section*{Acknowledgments}
This work was supported by the EPSRC. We acknowledge useful discussions with Prof. Bryan Gallagher.

\section*{Author Contributions}
N.A.P. fabricated and measured the samples, analysed the data, and performed the numerical calculations. C.H.M. supervised the project. Both authors wrote the manuscript.

\section*{Additional Information}
The authors declare no competing financial interests. Supplementary information accompanies this paper on www.nature.com/naturematerials. Reprints and permissions information is available online at http://npg.nature.com/reprintsandpermissions. Correspondence and requests for materials should be addressed to C.H.M.

\end{document}